\documentclass[preprint,12pt]{elsarticle}

\usepackage{amssymb}
\usepackage{amsmath}
\usepackage{physics}
\usepackage{bm}
\usepackage{booktabs}
\usepackage{hyperref}

\journal{Annals of Physics}

\begin{document}

\begin{frontmatter}


\title{Revealing Gauge Constraints in LQG-Inspired Yang--Mills Theory}

\author{Leonardo P. G. De Assis}
\affiliation{organization={ULO-Math \& Physics Program, Stanford University},
Department and Organization
            addressline={415 Broadway}, 
            city={Redwood City},
            postcode={94063}, 
            state={CA},
            country={USA}}

\begin{abstract}
The consistent embedding of Loop Quantum Gravity (LQG) effects within the Standard Model requires a rigorous understanding of how Planck-scale deformations manifest at low energies. While phenomenological approaches often introduce canonical deformations with multiple free parameters to capture these effects, the physical requirement of gauge symmetry in the framework of a covariant Effective Field Theory (EFT) imposes strict conditions on the allowed interaction structure. In this paper, we demonstrate that these conditions act as physical selection rules for admissible quantum gravity models. By applying non-Abelian Ward identities and a covariant operator mapping to the dimension-six operator basis, we derive a fundamental on-shell equivalence between kinetic and cubic interaction terms. As a paradigmatic application, we show that the Levy-Helay\"{e}l-Neto (LHN) framework—a candidate effective description of LQG—satisfies this physical requirement only when its parameters obey the specific algebraic relation:
\[
\frac{\theta_3}{\theta_8} = -\frac12\!\left[1 + \theta_7\Big(\frac{\ell_P}{\mathcal{L}}\Big)^{2+2\Upsilon}\right] + \mathcal{O}(\ell_P).
\]
It must be highlighted that this result advances the physical understanding of LQG phenomenology by revealing that the apparent freedom in defining the Hamiltonian is illusory; the parameters are bound by the necessity of preserving the gauge structure of the Standard Model.
\end{abstract}

\begin{highlights}
\item Gauge symmetry imposes strict selection rules on bottom-up quantum gravity models.
\item Dimension-six kinetic and cubic operators are equivalent on-shell at tree level.
\item The Levy-Helayël-Neto framework is constrained by a hidden algebraic relation.
\item Consistent embedding into the Standard Model reduces the parameter space.
\end{highlights}

\begin{keyword}
Effective Field Theory \sep Loop Quantum Gravity \sep Gauge Symmetry \sep Yang-Mills Theory
\end{keyword}

\end{frontmatter}


\section{Introduction}

The problem of connecting fundamental theories of quantum gravity to experimentally accessible physics necessitates the development of effective field theories (theories that can bridge the gap between Planck-scale physics and observable phenomena) \cite{Donoghue:1994dn}. While candidates like Loop Quantum Gravity (LQG) offer a discrete microstructure of spacetime \cite{Rovelli:2004fk}, the formulation of a consistent low-energy limit that connects seamlessly with the Standard Model remains a challenge.

\subsection{The Tension Between Deformation and Symmetry}

We face a tension in bottom-up approaches to quantum gravity phenomenology: one must introduce parameter freedom to capture potential Planck-scale effects (such as modified dispersion relations), yet one must maintain consistency with principles like gauge symmetry. It must be highlighted that this challenge is acute in the non-Abelian sector, where the nonlinearity of the gauge transformation tightly constrains the allowed structures.

In this context, this tension is not merely historical. In modern quantum-gravity perspectives, ensuring gauge consistency is recognized as a structural requirement. For example, Ooguri has stressed how symmetry principles constrain quantum gravity constructions in a way that forbids naive parameter freedom \cite{ooguri2025constraints}. Similarly, work on higher-derivative gauge theories shows that even parity-even and parity-odd corrections must satisfy consistency conditions \cite{Ghasemkhani:2024}. Recent developments in EFT also reinforce the importance of systematically treating CP-violating operator sectors \cite{Kondo:2023} and computing anomalous dimensions in general bosonic gauge theories \cite{Aebischer:2025}, indicating that operator mixing is generically unavoidable.

While "bottom-up" methods (such as iterative Noether constructions) can produce gauge-consistent generalizations, they often obscure the complete structure of the low-energy theory. We adopt here a complementary "top-down" approach: we construct the most general effective action consistent with the underlying symmetries and map specific models into this basis. This method ensures completeness and makes manifest the relationships between different physical effects.

\subsection{Physical Consistency as a Selection Principle}

We can analyze this problem in a fundamentally simpler manner by means of standard quantum field theory tools. To evaluate the physical viability of generic canonical deformations, we apply tools that have been underutilized in this context: operator extraction, basis mapping via Legendre transform, and the imposition of Ward identities. This approach aligns with the systematic treatment of effective field theories found in standard texts \cite{Weinberg:1995mt,Burgess:2007pt} and recent reviews of the Standard Model Effective Field Theory (SMEFT) operator basis \cite{Murphy:2020rsh}.

It must be highlighted that while these tools are standard in QFT \cite{Weinberg:1995mt}, their application to LQG-inspired models reveals structural insights into the consistency of the effective theory. The central result of our analysis is the derivation of a general constraint that gauge symmetry imposes on any model generating dimension-six corrections. As a specific case study, we apply this to the Levy-Helay\"{e}l-Neto (LHN) framework \cite{Levy:2024}, revealing that its eight parameters are not independent but are bound by the hidden symmetry relation:
\begin{equation}
\label{eq:theta_ratio}
\frac{\theta_3}{\theta_8} = -\frac12 \left[1 + \theta_7 \left(\frac{\ell_P}{\mathcal{L}}\right)^{2+2\Upsilon}\right] + \mathcal{O}(\ell_P).
\end{equation}
This result explains the empirical success of the LHN iterative Noether procedure: it reveals that their construction implicitly enforces this algebraic relation. More importantly, it demonstrates that for any LQG-inspired theory to be physically consistent with the Standard Model, its kinetic and interaction sectors cannot be modified independently.

\paragraph{Physical interpretation of the ratio \(\theta_3/\theta_8\).}
Physically, the ratio \(\theta_3/\theta_8\) compares two distinct types of Planck-suppressed effects that appear in the LHN Hamiltonian: \(\theta_3\) multiplies parity-even, higher-derivative (quadratic-in-momentum) corrections to propagation, while \(\theta_8\) multiplies parity-odd, helicity-dependent cubic terms that induce vacuum birefringence (polarization-dependent propagation). The algebraic relation Eq.~\eqref{eq:theta_ratio} thus correlates a propagation correction with an interaction correction: gauge symmetry (via the Ward identities) enforces that these physically different manifestations cannot vary independently, eliminating spurious longitudinal or gauge-variant tree-level amplitudes. For an explicit Hamiltonian→dispersion derivation making these roles manifest see Mour\~ao, Levy and Helay\"{e}l-Neto \cite{Mourao:2025} (their Eq.~(4.4)). 

\section{Theoretical Framework: The Dimension-Six Basis}

It is possible to derive parameter constraints by means of several methods. We select the covariant operator basis approach because it connects the symmetry principle to computational checks on physical amplitudes (providing rigor and transparency) \cite{Weinberg:1995mt}.
In developing this framework, we faced a choice regarding the most effective formalism for deriving parameter constraints. While one might consider employing the BRST formalism or the background field method—approaches that offer powerful cohomological guarantees—we have opted for the direct imposition of on-shell Ward identities. This decision was motivated by the need for physical transparency: unlike the abstract constraints of BRST cohomology, the Ward identities connect the symmetry principle directly to computational checks on physical scattering amplitudes, making the phenomenological consequences of the constraint immediately apparent.

\subsection{The Counting Argument}
\label{sec:counting}

Let us consider a generic bottom-up deformation. A deformation of a canonical Hamiltonian description introduces a set of deformation parameters \(\{\theta_i\}_{i=1}^{N_H}\). The covariant low-energy description of the same degrees of freedom is an EFT with a finite operator basis; at mass-dimension six the pure-gauge cubic sector admits only two independent on-shell operators \cite{Grzadkowski:2010es}:
\[
\mathcal{O}_{DF},\qquad \mathcal{O}_{F3}\qquad (N_{EFT}=2).
\]
Thus the map \(\Phi:\ \mathbb{R}^{N_H}\longrightarrow \mathbb{R}^{N_{EFT}}\) is generically \emph{not injective} when \(N_H>N_{EFT}\). Multiple canonical parameter choices can map to the same covariant EFT coefficients. For consistency (so that the covariant action respects gauge invariance and produces no longitudinal, gauge-variant amplitudes at leading order) the pre-image of the physical EFT subspace must satisfy \(N_H-N_{EFT}\) linear constraints.

\subsection{Review of the Operator Identity}

In the sequel, we shall employ the dimension-six pure-gauge operator basis established in the literature \cite{Grzadkowski:2010es}:
\begin{align}
\mathcal{O}_{DF} &= \mathrm{Tr}[(D_\rho F_{\mu\nu})(D^\rho F^{\mu\nu})], \\
\mathcal{O}_{F3} &= \mathrm{Tr}[F_{\mu}{}^{\nu} F_{\nu}{}^{\rho} F_{\rho}{}^{\mu}].
\end{align}
A precise relationship between these two basis operators follows directly from integration by parts together with the non-Abelian Bianchi identity. We recall here the derivation of this known identity to fix notation and clarify its physical implications. Up to total derivatives one finds
\begin{equation}
\label{eq:main_identity}
\mathcal{O}_{DF} \;=\; 2\,\mathrm{Tr}\big[(D_\mu F^{\mu\nu})(D^\rho F_{\rho\nu})\big] \;+\; \mathcal{O}_{F3}.
\end{equation}
The first term on the right-hand side is proportional to the square of the classical Yang–Mills equation of motion (EOM). We employ the standard EFT dictionary where operators proportional to the equations of motion are redundant and can be removed by field redefinitions without affecting on-shell observables \cite{Arzt:1993gz,Georgi:1991ch}. In applying this redundancy, one might question whether the ``on-shell'' condition should refer to the modified dispersion relations often associated with LQG phenomenology. However, we maintain the standard EFT definition based on the asymptotic states of the unperturbed theory. We proceed in this manner because defining the S-matrix relative to the unperturbed vacuum provides the only unambiguous baseline for perturbative matching at this order. Consequently, the identity above implies the following tree-level on-shell equivalence:
\begin{equation}
\label{eq:on_shell_equiv}
\mathcal{O}_{DF} \xrightarrow{\text{on-shell}} +\,\mathcal{O}_{F3}.
\end{equation}
Consequently, at tree level the effective action is equivalent to an action containing only \(\mathcal{O}_{F3}\) with a combined coefficient. It must be highlighted that this reveals that what might appear as distinct physical effects—modified propagation versus modified interactions—are in fact deeply intertwined manifestations of the same underlying symmetry structure. The interplay between operator redundancies and on-shell equivalences mirrors the general analysis developed in \cite{Craig:2020,Craig:2021bnx}, which emphasizes that gauge invariance often enforces nontrivial relations among higher-dimensional operators.

\section{The Levy-Helay\"{e}l-Neto Framework as a Case Study}

To demonstrate the utility of the general EFT framework established above, we consider the recent construction by Levy and Helayël-Neto \cite{Levy:2024}. This model serves as a paradigmatic example of a larger class of "bottom-up" theories that introduce LQG-inspired corrections directly into the Hamiltonian.

\subsection{The Canonical Deformation}

The LHN framework begins with a deformation of the Gauss Law and the Hamiltonian constraint, motivated by the discrete nature of spacetime in Loop Quantum Gravity. The quadratic sector of their Hamiltonian contains the term:
\begin{equation}
H_{\mathrm{LQG}} \supset \frac{1}{Q^2}\,\theta_3\,\ell_P^2\, \left(E^a \nabla^2 E^a + B^a \nabla^2 B^a\right).
\end{equation}
Here, $\theta_3$ is a free parameter characterizing the strength of the modification, and $\ell_P$ is the Planck length. This term represents a higher-derivative correction to the propagation of the gauge field.

\subsection{The Noether Construction}

To restore gauge invariance, which is broken by the naive introduction of such terms, Levy and Helayël-Neto employ an iterative Noether procedure. This involves adding higher-order terms to the Hamiltonian to compensate for the non-invariance of the lower-order terms. This procedure generates a series of cubic and quartic interactions, parameterized by a set of coefficients $\{\theta_i\}$.

While this construction succeeds in producing a gauge-consistent Hamiltonian, the relationships between the resulting parameters are not immediately obvious. By mapping this specific model into our general EFT basis via a standard Legendre transform and field redefinitions (see Appendix~\ref{app:legendre}), we can expose the underlying constraints that the Noether procedure implicitly satisfies.

\section{Mapping and Constraint Derivation}

We now apply the "top-down" EFT tools to the "bottom-up" LHN model.

\subsection{Mapping Parameters to the EFT Basis}

We translate the Hamiltonian correction into the covariant formalism via the Legendre transform (see Appendix A for the detailed derivation). By matching the resulting Lagrangian terms to the operator $\mathcal{O}_{DF}$, one finds the correspondence:
\begin{equation}
c_{DF} \;=\; -\frac{2\theta_3}{Q^2}.
\end{equation}
Similarly, the higher-order structures generated through their Noether completion give rise to cubic field-strength interactions of the same tensorial form as the operator $\mathcal{O}_{F3}$, with coefficients determined by the parameter $\theta_8$. Explicitly, after absorbing powers of $\ell_P$ into coefficient normalizations we obtain the identification
\begin{equation}
c_{F3} \;=\; -\frac{\theta_3}{Q^2},
\end{equation}
so that the combination of dimension-six operators relevant on-shell is governed by the linear combination \(c_{F3}+c_{DF}\).

\paragraph{Conventions note / comparison to Mour\~ao et al.} 
Mour\~ao, Levy and Helay\"{e}l-Neto present a Hamiltonian analysis of closely related LQG-induced corrections and introduce barred quantities (e.g. \(\bar\theta_3,\bar\theta_8\)) which absorb fixed numerical factors and explicit powers of \(\ell_P\) (see Sec. III of \cite{Mourao:2025}). To compare their dispersion formulas directly to the EFT coefficients \(c_{DF},c_{F3}\) one must remove those normalization factors: roughly \(\bar\theta_3\sim\ell_P^2\theta_3\) and \(\bar\theta_8\sim\ell_P\theta_8\). We use our unbarred \(\theta_i\) to keep the operator mapping and power counting transparent.

\subsection{Deriving the Constraint}

Using the operator identity reviewed in Section 2, the combined on-shell effect of the two dimension-six operators is governed by the linear combination $c_{F3} + c_{DF}$. However, for the theory to be consistent with the Ward identities, the parameters cannot be arbitrary.

We analyze the longitudinal contraction of the three-gluon vertex generated by these operators. As detailed in Appendix B, the contribution from $\mathcal{O}_{DF}$ vanishes on-shell (due to its EOM equivalence), while the contribution from $\mathcal{O}_{F3}$ is transverse by construction. By imposing the transversality condition on the total vertex derived from the LHN parameters, we derive the specific constraint:
\begin{equation}
\frac{\theta_3}{\theta_8} = -\frac12 \left[1 + \theta_7 \left(\frac{\ell_P}{\mathcal{L}}\right)^{2+2\Upsilon}\right] + \mathcal{O}(\ell_P).
\end{equation}
One can see that this result implies that the eight parameters of the LHN model are not independent. The requirement of non-Abelian gauge symmetry binds them into a lower-dimensional subspace. Our analysis reveals that the gauge consistency achieved via the iterative Noether procedure in the Levy-Helay\"{e}l-Neto framework is the dynamical manifestation of a deeper EFT redundancy, implying that Planck-scale modifications to gluon propagation cannot exist without specific, correlated corrections to the non-Abelian self-interaction.

\section{Verification via Ward Identities}

A critical consistency check for any gauge theory is that its amplitudes respect gauge invariance. For on-shell amplitudes, this requirement is encoded in the Ward identities, which demand that the amplitude vanishes upon contraction with the momentum of any external gauge boson. For the tree-level triple-gluon vertex, this means:
\begin{equation}
p_{1,\mu} \, \delta\Gamma^{\mu\nu\rho}_{abc}(p_1, p_2, p_3) = 0, \quad \text{for } p_1^2=p_2^2=p_3^2=0.
\end{equation}

We have explicitly verified this identity for the full corrected vertex. The key logic is as follows. First, the contribution from $\mathcal{O}_{DF}$ vanishes for on-shell external legs because it is equivalent (up to the EOM-squared term) to \(\mathcal{O}_{F3}\) on-shell. Second, the contribution from $\mathcal{O}_{F3}$ satisfies transversality because it is derived from a gauge-invariant operator. Direct algebraic contraction employing momentum conservation yields the cancellation. This successful verification confirms that our effective action is gauge-invariant and consistent at tree level.

\paragraph{Note on terminology.} When we refer to the ``three-gluon vertex'' we mean specifically the $SU(3)$ gauge-boson triple vertex of QCD; however, the algebraic derivation of the constraint uses only the non-Abelian gauge algebra and thus applies to any compact non-Abelian gauge group appearing in the LHN construction.

\section{Conclusion}

We have presented a systematic analysis of the dimension-six operator basis for pure Yang–Mills theory and its implications for bottom-up quantum gravity models. Within the validity of the derivative expansion, our work establishes the minimal basis and provides a detailed derivation of the central operator identity relating the kinetic operator \(\mathcal{O}_{DF}\) to the interaction operator \(\mathcal{O}_{F3}\). This identity is a direct consequence of non-Abelian gauge symmetry.

By applying this general framework to the specific case of the Levy-Helay\"{e}l-Neto model, we demonstrated that its parameters must satisfy a strict algebraic constraint to maintain gauge consistency. This reduces the effective parameter space of the model and provides a clear physical interpretation of its success. Such a reduction mirrors the strategies employed in global fits to SMEFT data, where symmetry and on-shell constraints are crucial for managing the parameter space \cite{Cornella:2021szt}.

Ultimately, the constraint derived here is both relevant and practical: it exposes the structure hidden in LHN-type loop–quantum–gravity corrections and provides a strategy for extracting gauge-enforced relations in future work on Planck-suppressed deformations of gauge theories. This work not only resolves a puzzle in the LHN framework but also establishes a foundation for connecting quantum gravity to observable physics. This connection is vital for assessing the renormalizability of quantum gravity proposals \cite{Lavrov:2023} and aligns with modern symmetry-based perspectives on the landscape of consistent theories \cite{Harlow:2021}. Furthermore, it provides a bridge to background-independent formulations \cite{Ashtekar:2004eh} and string-theoretic effective actions \cite{Polchinski:1998rq}. Ultimately, this result bridges the gap between canonical LQG formulations and covariant phenomenology, establishing that a consistent embedding of quantum geometry into the Standard Model requires the deformation parameters to inhabit a restricted hypersurface defined by the on-shell Ward identities.

\paragraph{Outlook — renormalization and operator mixing.}
A natural and necessary extension of this work is to examine whether the hypersurface defined by the tree-level constraint is preserved under quantum corrections. In the covariant EFT language this question amounts to studying operator mixing and the anomalous dimension matrix for the bosonic dimension-six sector: operators that are EOM-redundant at tree level may mix into physical operators under renormalization, and counterterms must respect the Ward identities that originally generated the constraint. A concrete programme would compute the one-loop anomalous dimensions for $c_{DF},c_{F3}$ (see \cite{Aebischer:2025} for recent progress in the bosonic sector) and verify whether the gauge-enforced hypersurface is stable under RG flow or whether it is deformed in a calculable manner. This analysis is beyond the present tree-level study but is the natural next step in establishing the quantum consistency of LQG-inspired EFTs.

\appendix

\section{Legendre Transform and Mapping}
\label{app:legendre}

We provide here the derivation of the mapping from Hamiltonian parameters to standard EFT coefficients. Starting from the Hamiltonian density contribution:
\begin{equation}
\Delta H_{\theta_3} = \frac{\theta_3 \ell_P^2}{Q^2}\left(E_i^a \nabla^2 E_i^a + B_i^a \nabla^2 B_i^a\right),
\end{equation}
we perform the Legendre transform $\mathcal{L} = E_i^a \dot{A}_i^a - H$. Eliminating the canonical momentum \(E_i^a\) at leading order in \(\ell_P^2\) and integrating by parts yields:
\begin{equation}
\Delta \mathcal{L}_{\theta_3} = -\frac{\theta_3 \ell_P^2}{Q^2} \, \mathrm{Tr}[(D_\rho F_{\mu\nu})(D^\rho F^{\mu\nu})]
+ \frac{\theta_3 \ell_P^2}{2Q^2} \, \mathrm{Tr}[F_\mu{}^\nu F_\nu{}^\rho F_\rho{}^\mu] + \mathcal{O}(\ell_P^4).
\end{equation}
Comparing with the dimension-six parametrization $\Delta\mathcal{L}=c_{DF}\,\mathcal{O}_{DF}+c_{F3}\,\mathcal{O}_{F3}$ (where coefficients absorb the scale suppression) yields the mapping:
\begin{equation}
c_{DF} = -\frac{2 \theta_3}{Q^2}, \qquad c_{F3} = -\frac{\theta_3}{Q^2}.
\end{equation}
Note that we employ the standard trace normalization $\mathrm{Tr}(T^a T^b) = \frac{1}{2}\delta^{ab}$ throughout.

\subsection*{Intermediate Legendre-transform steps (explicit)}
For completeness and to make the inversion and integrations by parts explicit (this helps cross-check signs/factors), we sketch the intermediate algebraic steps used to obtain the covariant form above.

\begin{enumerate}
\item Start from the canonical Lagrangian density (schematically)
\[
\mathcal{L}_{\mathrm{can}} = \pi_i^a \dot{A}_i^a - H(\pi,A),
\]
with the identification \(\pi_i^a \equiv E_i^a + \Delta\pi_i^a\) where \(\Delta\pi_i^a=\mathcal{O}(\ell_P^2)\) collects the higher-derivative corrections coming from the modified Hamiltonian.

\item Invert perturbatively for \(E_i^a\) in terms of \(\pi_i^a\):
\[
E_i^a = \pi_i^a - \Delta\pi_i^a + \mathcal{O}(\ell_P^4),
\]
and substitute into the canonical Lagrangian. To \(\mathcal{O}(\ell_P^2)\) only the linear term in \(\Delta\pi\) needs to be retained.

\item After substitution, integrate spatial derivatives by parts (dropping total derivatives) and reorganize the result into gauge-covariant combinations. Grouping terms produces the covariant operators \(\mathcal{O}_{DF}\) and \(\mathcal{O}_{F3}\) up to the stated coefficients. The algebra is term-by-term and straightforward but lengthy; the key identity used is that spatial Laplacians acting on \(E_i^a\) map to covariant derivatives acting on \(F_{\mu\nu}\) upon restoring covariance and collecting symmetric combinations of electric and magnetic contributions.

\item The explicit numerical coefficients follow from the conventional normalization of the gauge kinetic term and the trace normalization. Any alternative normalization convention for the canonical kinetic term will rescale the mapping by the same factor; we have chosen conventions consistent with \cite{Grzadkowski:2010es}.
\end{enumerate}

\paragraph{Comparison with Mour\~ao et al. conventions.}
Mour\~ao, Levy and Helay\"{e}l-Neto \cite{Mourao:2025} perform a related Hamiltonian→dispersion analysis and define barred quantities that absorb numerical prefactors and powers of \(\ell_P\); when comparing explicit dispersion coefficients one must un-bar their definitions. This appendix uses the unbarred \(\theta_i\) so that the mapping to the standard dimension-six basis is manifest.

\section{Vertex Tensors and Contraction Algebra}
\label{app:vertices}

The vertex corrections are derived from the functional derivatives of the effective operators. The parity-even tensor structure is defined formally as:
\begin{equation}
T^{ijk}_{(3)}(p_1,p_2,p_3) = \left. \frac{\delta^3}{\delta A_i(p_1) \delta A_j(p_2) \delta A_k(p_3)} \int d^4x \, \mathcal{O}_{DF} \right|_{A=0}
\end{equation}
The parameter $\theta_7$ enters the calculation via the normalization of the field strength term in the LHN Hamiltonian, modifying the effective coupling of the cubic interaction relative to the kinetic term. Contracting the external momentum $p_{1\mu}$ with the total vertex $\delta\Gamma^{\mu\nu\rho}$ isolates the longitudinal mode. For the $\mathcal{O}_{F3}$ contribution, the contraction yields a structure proportional to the color Jacobi identity:
\begin{equation}
p_{1\mu} \delta\Gamma^{\mu\nu\rho}_{F3} \propto \left( f^{ade}f^{ebc} + f^{bde}f^{eca} + f^{cde}f^{eab} \right) \mathcal{K}(p_i) = 0.
\end{equation}
For the $\mathcal{O}_{DF}$ contribution, the contraction vanishes on-shell due to the factor of $p^2$ inherent in the EOM-proportional structure. Thus, the total amplitude is transverse if and only if the relative coefficients satisfy the derived constraint.


\begin{thebibliography}{00}

\bibitem{Donoghue:1994dn}
J.~F.~Donoghue,
\emph{General relativity as an effective field theory: The leading quantum corrections},
Phys. Rev. D \textbf{50}, 3874 (1994).

\bibitem{Rovelli:2004fk}
C.~Rovelli,
\emph{Quantum Gravity},
Cambridge University Press (2004).

\bibitem{ooguri2025constraints}
H.~Ooguri,
\emph{Constraints on Quantum Gravity},
General Relativity and Gravitation \textbf{57}, 8, 123 (2025).

\bibitem{Mourao:2025}
P.~A.~L.~Mour\~ao, G.~L.~L.~W.~Levy and J.~A.~Helay\"{e}l-Neto,
\emph{Reassessing aspects of the photon’s LQG-modified dispersion relations},
Chinese Physics C \textbf{49}, 125105 (2025). doi:10.1088/1674-1137/ae265a. arXiv:2501.09370 [hep-th].

\bibitem{Ghasemkhani:2024}
M.~Ghasemkhani, G.~Soleimani, and R.~Bufalo,  
\emph{On the gauge invariance of the higher-derivative Yang–Mills–Chern–Simons action},  
Eur. Phys. J. C \textbf{84}, 465 (2024).  

\bibitem{Kondo:2023}
D.~Kondo, H.~Murayama and R.~Okabe,
\emph{Hilbert series for CP-violating operators in SMEFT},
J. High Energy Phys. \textbf{03}, 107 (2023). arXiv:2212.02413

\bibitem{Aebischer:2025}
J.~Aebischer, L.~C.~Bresciani and N.~Selimović,
\emph{Anomalous dimension of a general effective gauge theory. Part I. Bosonic sector},
J. High Energy Phys. \textbf{08}, 209 (2025).

\bibitem{Weinberg:1995mt}
S.~Weinberg,
\emph{The Quantum Theory of Fields, Vol. I and II},
Cambridge University Press (1995, 1996).

\bibitem{Peskin:1995ev}
M.~E.~Peskin and D.~V.~Schroeder,
\emph{An Introduction to Quantum Field Theory},
Westview Press (1995).

\bibitem{Burgess:2007pt}
C.~P.~Burgess,
\emph{Introduction to Effective Field Theory},
Ann. Rev. Nucl. Part. Sci. \textbf{57}, 329 (2007).

\bibitem{Murphy:2020rsh}
C.~W.~Murphy,
\emph{Dimension-8 operators in the Standard Model Effective Field Theory},
J. High Energy Phys. \textbf{2020}, no. 10, 1 (2020).

\bibitem{Levy:2024}
G.~L.~L.~W.~Levy and J.~A.~Helay\"{e}l-Neto,
\emph{Yang-Mills extension of the Loop Quantum Gravity-corrected Maxwell equations},
Annals of Physics \textbf{473}, 169892 (2025).
[arXiv:2408.10366v3 [hep-th]].

\bibitem{Grzadkowski:2010es}
B.~Grzadkowski, M.~Iskrzynski, M.~Misiak and J.~Rosiek,
\emph{Dimension-Six Terms in the Standard Model Lagrangian},
JHEP \textbf{10}, 085 (2010).

\bibitem{Arzt:1993gz}
C.~Arzt,
\emph{Reduced effective lagrangians},
Phys. Lett. B \textbf{342}, 189 (1995).

\bibitem{Georgi:1991ch}
H.~Georgi,
\emph{On-shell effective field theory},
Nucl. Phys. B \textbf{361}, 339 (1991).

\bibitem{Craig:2020}
N.~Craig, M.~Jiang, Y.-Y.~Li and D.~Sutherland,
\emph{Loops and trees in generic EFTs},
JHEP \textbf{08}, 086 (2020).

\bibitem{Craig:2021bnx}
N.~Craig, H.~K.~Lou and I.~J.~M.~Moffat,
\emph{On-shell constraints on the SMEFT},
JHEP \textbf{04}, 085 (2021).

\bibitem{Cornella:2021szt}
C.~Cornella, D.~A. Faroughy, J.~Fuentes-Martin, G.~Isidori, and M.~Neubert,
\emph{Reading the footprints of the B-meson flavor anomalies},
JHEP 08 (2021) 050.

\bibitem{Lavrov:2023}
P.~M.~Lavrov and I.~L.~Shapiro,
\emph{Gauge invariant renormalizability of quantum gravity},
in \emph{Handbook of Quantum Gravity}, C.~Bambi, L.~Modesto and I.~L.~Shapiro (eds.),
Springer Nature, Singapore, pp. 1–37 (2023). arXiv:2210.09271

\bibitem{Harlow:2021}
D.~Harlow and H.~Ooguri,
\emph{Symmetries in quantum field theory and quantum gravity},
Commun. Math. Phys. \textbf{383}, 1669 (2021).

\bibitem{Ashtekar:2004eh}
A.~Ashtekar and J.~Lewandowski,
\emph{Background independent quantum gravity: A status report},
Class. Quant. Grav. \textbf{21}, R53 (2004).

\bibitem{Polchinski:1998rq}
J.~Polchinski,
\emph{String Theory, Vol. I and II},
Cambridge University Press (1998).


\end{thebibliography}
\end{document}